\documentclass[11pt]{article}
\title{Chapter 6\\Toward a Consistent Theory of Relativistic Rotation\\ in \textit{Relativity in Rotating Frames} (Kluwer Academic)}
\author{Robert D. Klauber\\1100 University Manor Dr., 38B, Fairfield, IA 52556, USA\\klauber@.iowatelecom.net or rklauber@.netscape.net\cite{email}}
\date{January 2004}

\usepackage{graphicx}
\usepackage{longtable}

\begin{document}

\maketitle

\begin{abstract}

\begin{center}
\textbf{Part 1: Traditional Analysis Conundrums}
     
\end{center}

Although most physicists presume the theoretical basis of relativistically 
rotating systems is well established, there may be grounds to call the 
traditional analysis of such systems into question. That analysis is argued 
to be inconsistent with regard to its prediction for circumferential Lorentz 
contraction, and via the hypothesis of locality, the postulates of special 
relativity. It is also contended that the traditional analysis is in 
violation of the continuous and single valued nature of physical time. It is 
further submitted to be in disagreement with the empirical finding of 
Brillet and Hall, the global positioning system satellite data, and a light 
pulse arrival time analysis of the Sagnac experiment. 

\begin{center}
\textbf{Part 2: Resolution of the Conundrums:\\Differential Geometry and Non-time-orthogonality}
\end{center}

It is postulated that physical constraints on time (its continuous and 
single-valued nature) limit the set of possible synchronization/simultaneity 
schemes in rotation to one, the ``flash from center'' scheme. A 
differential geometry analysis based on this simultaneity postulate is 
presented in which the rotating frame metric is constrained to be locally 
non-time-orthogonal (NTO) and due to which, all inconsistencies and 
disagreements with experiment are resolved. The hypothesis of locality is 
shown to be invalid for rotation specifically, and generally valid only for 
non-inertial frames in which the metric can have all null off diagonal 
space-time components (i.e., time is orthogonal to space.) The analysis 
approach presented does not contravene traditional relativistic theory for 
translating systems and makes many (but not all) of the same predictions for 
rotating systems as does the traditional (time orthogonal) analysis.

\begin{center}
\textbf{Part 3: Experiment and Non-time-orthogonal Analysis}
\end{center}

Experiments performed from the 1880s to the present to test special 
relativity are summarized, and their relevance to NTO analysis is presented. 
One test, that of Brillet and Hall, appears capable of discerning between 
the NTO and traditional approaches to relativistic rotation. It yielded a 
signal predicted by NTO analysis, but not by the traditional approach. Other 
evidence in favor of the NTO approach may be inherent in the global 
positioning system data, and the Sagnac experiment.

%%after abstract, insert following
\end{abstract}

\section{Traditional Analysis Conundrums}
\subsection{Introduction}
Part 1 outlines the traditional approach to relativistic rotation and 
discusses various apparent inconsistencies associated therewith. Following 
an analysis of synchronization/simultaneity in rotating frames and seeming 
traditional approach problems therein, a new postulate is introduced, which 
will be used in Part 2 to pose an alternative approach to resolving the 
inconsistencies.

\subsection{Relevant Relativity Principles}
\label{subsec:relevant}
Special relativity theory (SRT) is restricted to inertial systems and is 
derived from two symmetry postulates:

\begin{enumerate}
\item The speed of light is the same for all inertial observers (it is invariant) and equals $c.$
\item There is no preferred inertial reference frame. (Velocity is relative, and the laws of nature are covariant, i.e., the same for all inertial observers.)
\end{enumerate}

The first postulate, applied to the one-way speed of light, is equivalent to 
demanding that Einstein synchronization of clocks holds. In Einstein 
synchronization, one starts from a first clock at time $t_{A}$ on that clock 
and sends a light pulse to a second clock fixed in the same frame as the 
first. The light pulse is reflected back at the second clock and returns to 
the first clock at time $t_{B}$ on the first clock. The time on the second 
clock is then set such that its reading when the light was reflected would 
have been ($t_{A}+t_{B})$/2, the time on the first clock half way between 
the emission and reception times. This ensures the one-way speed of light, 
measured as the distance traveled between clocks divided by the time 
difference of the two clocks, is always $c$.

In recent years, many relativists have come to consider Einstein 
synchronization merely a convention, or gauge, that affects no measurable 
quantities\cite{Anderson:1998}$^{,}$\cite{Minguzzi:2002}. For example, in 
all such gauge theories of synchronization, the round trip speed of light is 
$c$ (though the one-way speed of light need not be.) Nevertheless, underlying 
SRT is the assumption that Einstein synchronization is always one of the 
possible conventions that makes valid predictions about inertial frames in 
the physical world.

General relativity is applicable to non-inertial systems and is based on 
additional postulates, including the equivalence principle and the 
hypothesis of locality (or sometimes, the ``surrogate frames postulate''). 
The hypothesis of locality stands as a linchpin in the traditional approach 
to relativistic rotation, and thus, I number it among the postulates of 
importance to this article.

\begin{enumerate}
\setcounter{enumi}{2}
\item Hypothesis of locality: Locally (i.e., over infinitesimal regions of space and time), neither gravity nor acceleration changes the length of a standard rod or the rate of a standard clock relative to a nearby freely falling (i.e., inertial) standard rod or standard clock instantaneously co-moving with it. See M{\o}ller\cite{ller:1969}, Einstein\cite{Stachel:1950}, and Mashoon\cite{Mashoon:2003}.
\end{enumerate}

Stated another way, a local inertial observer is equivalent to a local 
co-moving non-inertial observer in all matters having to do with 
measurements of distance and time. It follows immediately that Einstein 
synchronization can be carried out locally, and that for such 
synchronization, the local one-way speed of light measured in a 
\textit{non-inertial} frame is $c$. Hence, a Lorentz frame can be used as a local surrogate for the 
non-inertial frame. This has a basis in differential geometry, in which a 
curved space is locally flat and can be represented locally by Cartesian 
coordinates.

Minguzzi\cite{Minguzzi:1} and M{\o}ller\cite{See:1}, among others, note 
that the hypothesis of locality is only an assumption. It is, however, an 
assumption that, historically, has worked very well in a large number of 
applications. See, for example, the treatment of acceleration by Misner, 
Thorne, and Wheeler\cite{Misner:1973} using instantaneous local Lorentz 
frames.

\subsection{The Traditional Approach}
\label{subsec:mylabel1}
The traditional approach to relativistic rotation assumes the hypothesis of 
locality is a fundamental and universal truth. As done successfully in 
other, non-rotating, cases, values in local co-moving Lorentz frames are 
integrated to determine global values for quantities such as distance and 
time, which would, in principle, be measured with standard meter sticks and 
clocks by an observer in the rotating frame.

The oft-cited example, first delineated by Einstein\cite{Ref:1}, is the 
purported Lorentz contraction of the rim of a rotating disk. (Or 
alternatively, the circumferential stresses induced in the disk when the rim 
tries to contract but is restricted from doing so via elastic forces in the 
disk material.) A local Lorentz frame instantaneously co-moving with a point 
on the rim, it is argued, exhibits Lorentz contraction of its meter sticks 
in the direction of the rim tangent, via its velocity, \textit{v = $\omega $r}. This infinitesimal 
length contraction is subsequently integrated over all of the local Lorentz 
frames instantaneously at rest with respect to each successive point along 
the rim. The result is a number of meter sticks that is greater than 2$\pi 
r$, and thus, the disk surface is concluded to be non-Euclidean, or Riemann 
curved\cite{Transforming:1}$^{,}$\cite{Relativistic:2003}.

\subsection{Inconsistency of Circumferential Lorentz Contraction}
\label{subsec:inconsistency}
According to SRT, an observer does not see his own lengths contracting. Only 
a second observer moving relative to him sees the first observer's length 
dimension contracted. Hence, from the point of view of the disk observer, 
her own meter sticks are not contracted\cite{Some:1}, and there can be no 
curvature of the rotating disk surface. The traditional analysis is thus, 
inconsistent\cite{Tartaglia:1999}.

Consider further the disk observer looking out at the meter sticks at rest 
in the lab close to the disk's rim. Via the hypothesis of locality (in which 
she is equivalent to a local co-moving Lorentz observer), she sees the lab 
meter sticks as having a velocity with respect to her. Hence, by the 
traditional logic, she sees them as contracted in the circumferential 
direction. She must therefore conclude that the lab surface is curved. But 
those of us living in the lab know this is simply not true, and again the 
analysis is inconsistent.

Although these arguments seem to be rarely considered by traditionalists, 
when they are brought to their attention, the usual defense is that ``the 
rotating frame is not an inertial frame and thus is different''. Yet, the 
hypothesis of locality, the starting point for the analysis, assumes that 
they are not different in this regard.

Furthermore, if the non-inertial argument has any validity, then it must 
imply that the length contraction of the rim is absolute, i.e., both the lab 
and disk observer agree that the disk meter sticks are contracted. Yet, 
consider the limit case of low $\omega $, high $r$, such that $a=\omega 
^2r\approx 0$, while $v=\omega r$ is close to the speed of light (the 
``limit case''). Advocates of the traditional approach contend that, since 
the limit case observer fixed to the rotating disk rim feels no inertial 
``force'', she becomes, effectively, a Lorentz observer. In this case, each 
of the lab and disk observers must see the other's meter sticks as 
contracted and their own as normal. Yet, the non-inertial argument started 
with the assumption that the disk observer's meter sticks contracted in an 
absolute way, agreed to by all observers\cite{Nikolic:2003}.

\textbf{Conclusion}: Length contraction applied via the traditional analysis 
to rotating systems appears self-contradictory.

\subsection{Second SRT Postulate Not Valid in Rotation}
Without looking outside, an observer on the rim of a rotating disk can 
determine her angular velocity $\omega $, using, for example, a Foucault 
pendulum. She can also use a spring mass system to measure $kx/m=\omega 
^2r$, and hence determine $r$, the distance to the center of rotation. (The 
Newtonian limit is used to simplify the example. The conclusion is also true 
for relativistic calculations.)

That is, contrary to the dictate of the second postulate, there are 
experiments an observer can perform locally from entirely within the 
rotating frame to determine her speed in an absolute sense. (To be precise, 
her speed with respect to the inertial frame in which her center of rotation 
is fixed.) Her velocity is not relative. Both the lab and the rim fixed 
observers determine the same value for it. With respect to circumferential 
speed, there is a preferred frame, and both observers agree it is the one 
where such speed is zero, i.e., the non-rotating lab frame.

\textbf{Conclusion}: The second relativity postulate does not appear to hold 
for rotating systems

\subsection{First SRT Postulate: Thought and Sagnac Experiments }
\subsubsection{Thought Experiment}
\label{subsubsec:thought}

Consider the following thought experiment (see Selleri\cite{Selleri:1997}) 
involving an observer fixed to the rotating disk of Figure 1 who measures 
the speed of light.

The observer shown has already laid meter sticks along the rim circumference 
and determined the distance around that circumference. As part of her 
experiment, she has also set up a cylindrical mirror, reflecting side facing 
inward, all around the circumference. She takes a clock with her and anchors 
herself to one spot on the disk rim. When her clock reads time $T$ = 0 (left 
side of Figure 1) he shines two short pulses of light tangent to the rim in 
opposite directions. The mirror will cause these light pulses to travel 
circular paths around the rim, one clockwise (cw) and one counterclockwise 
(ccw).

From the lab, we see the cw and ccw light pulses having the same speed $c$. 
However, as the pulses travel around the rim, the rim and the observer fixed 
to it move as well. Hence, a short time later, as illustrated in the right 
side of Figure 1, the cw light pulse has returned to the observer, whereas 
the ccw pulse has yet to do so. A little later (not shown) the ccw pulse 
will have caught up to the observer.

For the observer, from her perspective on the disk, both light rays travel 
the same number of meters around the circumference. But her experience and 
her clock readings tell her that the cw pulse took less time to travel the 
same distance around the circumference than the ccw pulse.

\begin{figure}
\includegraphics[bbllx=-0.5in,bblly=0.13in,bburx=4.06in,bbury=1.93in,scale=1.00]{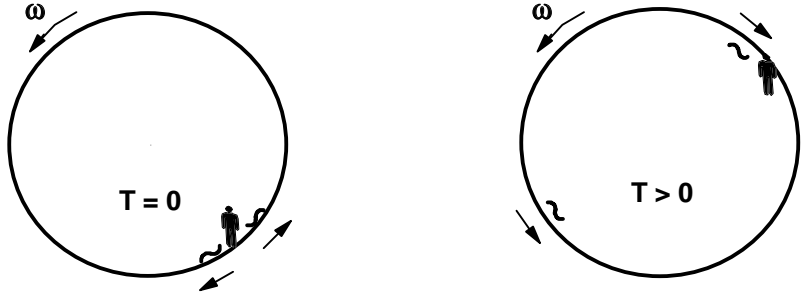}
\caption{Rotating Disk Observer Measuring Light Speed}
\label{fig1}
\end{figure}

She can only conclude that, from her point of view, the cw pulse traveled 
faster than the ccw pulse. Hence, the speed of light as measured on the 
rotating disk is anisotropic and different from that measured in the lab. 
Thus, one could conclude that the first relativity postulate, in the context 
of the hypothesis of locality, is violated.

Arguments against this conclusion are rooted in two interrelated concepts: 
i) purported differences between the global (as measured in the above 
thought experiment) and local, physical speeds of 
light\cite{For:1998}$^{,}$\cite{Also:2003}$^{,}$\cite{Also:2004}, and ii) 
the synchronization/simultaneity 
employed\cite{See:1998}$^{,}$\cite{Also:1}. The author has extensive 
remarks on this in a subsequent section, but for now, submits that the 
appropriate synchronization scheme comprises the following.

Consider the ccw light pulse and the time difference on the clock held by 
the observer in Figure 1 between the emission (assume initial clock time is 
$t_{A}$ = 0) and reception (clock time $t_{B})$ events. Employ the 
synchronization method of Section \ref{subsec:relevant}, only instead 
of using a back and forth round trip for light (Einstein synchronization), 
use a circular round trip. That is, the time on the clock half way along the 
round trip ccw path (at 180\r{ } of the disk here), at the instant the ccw 
light pulse was there, should be ($t_{A}+t_{B})$/2 = $t_{B}$/2. With this 
time, the ccw speed of light will be the same as that computed for the round 
trip, i.e., it will be less than $c$. Now synchronize the clock at 90\r{ } the 
same way. Assume its setting at the instant the light pulse passed was half 
that on the clock at 180\r{ } when the light passed that clock (or 
equivalently, $\raise.5ex\hbox{$\scriptstyle 1$}\kern-.1em/ 
\kern-.15em\lower.25ex\hbox{$\scriptstyle 4$} $ of $t_{B}$.) Doing this, one 
again finds the ccw speed of light to be the same value, which is $<c$. 
Repeat over smaller intervals until, in the limit, one finds the local speed 
of light to be the same, and therefore not equal to $c$.

The entire procedure can be repeated for the cw light pulse, and one will 
find the clocks at each location done via the cw and ccw methods are 
synchronized, i.e., they are the same clocks. One will also find that the 
local speeds of light are anisotropic and equal to the same values as the 
global ones determined using a single clock at the emission/reception point.

\textbf{Conclusion:} While it may appear that the local speed of light, 
being anisotropic, violates the first relativity postulate, there is still 
the possibility that Einstein synchronization may be valid locally (as one 
of the possible local synchronization schemes). This would mean that for 
such synchronization, the local speed of light would be $c$, and one could get 
correct results using the hypothesis of locality and local Lorentz frame 
analyses.

\textbf{Challenge to traditionalists:} The author does not believe this and 
challenges advocates of the traditional approach to begin with the 
assumption of local isotropic light speed, and without looking outside of 
the rotating frame, kinematically derive the result that the cw light pulse 
arrives back at the emission point before the ccw one.

\subsubsection{The Sagnac Experiment}
\label{subsubsec:mylabel1}
In 1913, G. Sagnac\cite{Sagnac:1914} carried out a now famous experiment, 
similar in many ways to the thought experiment of Figure 1. On a rotating 
platform, he split light from a single source into cw and ccw rays that 
traveled identical paths in opposite directions around the platform. He 
combined the returning rays to form a visible interference pattern, and 
found that the fringes shifted as the speed of rotation changed. A number of 
others\cite{Dufour:1942}$^{,}$\cite{See:1967} subsequently performed the 
same test with the same results.

If the speed of light were locally invariant and always equal to $c$, then 
speeding up or slowing of the rotation rate of the platform should not 
change the location of the fringes. However, the fringes do change with 
speed and once again we have a test (Sagnac) whereby we can determine a 
preferred frame, in seeming violation of the second relativity postulate and 
the hypothesis of locality.

Putative explanations for this in the context of the traditional approach 
hinge, once again, on synchronization/simultaneity and global vs. local 
arguments. These are addressed in the following section.

I do contend that the thought experiment of Figure 1 makes it clear that 
any explanation for the Sagnac experiment, from the point of view of the 
disk reference frame, must account for different \textit{arrival times} for the cw and ccw light 
pulses. Analyses based on Doppler shifts\cite{Dresden:1979} or DeBroglie 
momentum/wave length\cite{Mashoon:1998} changes are simply not sufficient 
to explain this.

The calculation of this arrival time difference, derived from the lab frame, 
is well known and is repeated for reference in the Appendix.

\subsection{Synchronization/Simultaneity in the Traditional Approach}
\label{subsec:synchronization}
\subsubsection{The Traditional Approach ``Time Gap''}
Consider the non-rotating (lab) frame as K; the rotating (disk) frame as k. 
Figure 2 depicts inertial measuring rods in inertial frames K$_{1}$ to 
K$_{8}$ with speeds $\omega r$ instantaneously at rest with respect to eight 
points on the rotating disk rim as shown. For practical reasons, only eight 
finite length rods are shown, and one can consider them as a symbolic 
representation of an infinite number of such rods of infinitesimal length. A 
and B are \textit{events} located in space at the endpoints of the K$_{1}$ rod which are 
simultaneous as seen from K$_{1}$; B and C are events located in space at 
the endpoints of the K$_{2}$ rod which are simultaneous in K$_{2}$; and so 
on for the other events C to J. A,B, ...J can be envisioned as flashes of 
light emitted by bulbs situated equidistantly around the disk rim.

p is a spatial (three-dimensional) point fixed to the disk at which both A 
and J occur. q is the spatial point on the disk at which B occurs. In 
principle, A, B, ... J, as well as p and q are located on the disk rim 
though they may not look so in Figure 2, since the co-moving rods shown are 
not infinitesimal in length.

\begin{figure}
\includegraphics[bbllx=-1.75in,bblly=0.13in,bburx=4.45in,bbury=2.51in,scale=.5] {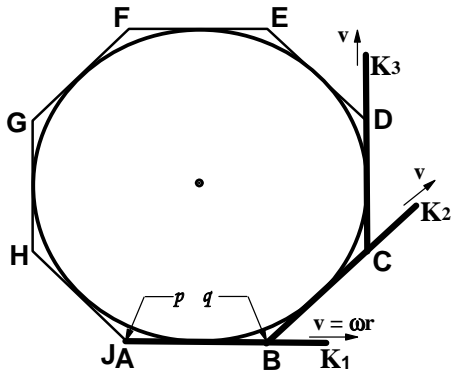}
\caption{Inertial Co-Moving Frames}
\label{fig2}
\end{figure}

\begin{figure}
\includegraphics[bbllx=-1.0in,bblly=1.13in,bburx=4.7in,bbury=3.51in,scale=.7] {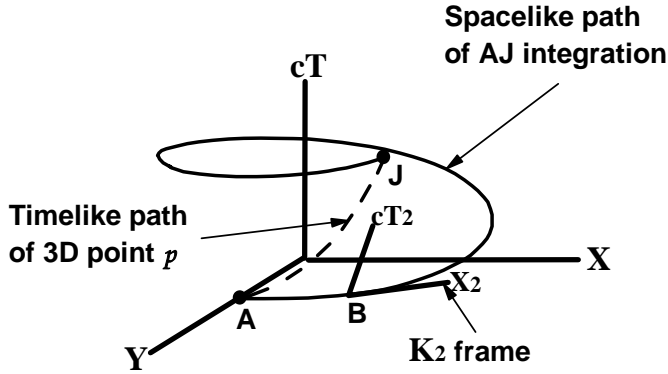}
\bigskip
\bigskip
\caption{Co-Moving Frames Integration Path}
\label{fig3}
\end{figure}

%\begin{figure}[t]
%\sidebyside {\includegraphics[width=5cm,height=4.5cm]{klaubfin2.eps}
%\caption{Inertial Co-Moving Frame} \label{fig:2}}
%{\includegraphics[width=5cm,height=4.5cm]{klaubfin3.eps}
%\caption{Co-Moving Frames Integration Path} \label{fig:3}}
%\end{figure}

In the traditional analysis, the hypothesis of locality is invoked to claim 
that times and distances measured by standard measuring rods and clocks in 
the local co-moving inertial frames are identical to those riding with the 
disk.

Note that although events A and B are simultaneous as seen from K$_{1}$, 
they are not simultaneous as seen in K (via SRT for two inertial frames in 
relative motion). As seen from K, A occurs before B. Similarly, B occurs 
before C, and so on around the rim. If the events are light flashes, a 
ground based observer looking down on the disk would see the A flash, then 
B, then C, etc. Hence we conclude that as seen from K, A occurs before J 
even though A and J are both located at the same 3D point p fixed to the 
rim. As seen from K, during the time interval between events A and J, the 
disk rotates, and hence the point p moves. (As an aside, Figure 2 can now be 
seen to be merely symbolic since events A to J would not in actuality be 
seen from K to occur at the locations shown in Figure 2. That is, by the 
time the K observer sees the B flash, the disk has rotated a little. It 
rotates a little more before he sees the C flash, etc.) 

According to the traditional treatment of the rotating disk, one then uses 
the K$_{i}$ rods and integrates (adds the rod lengths) along the path AB 
...J, moving sequentially from co-moving inertial frame to co-moving 
inertial frame. This path is represented by the solid line in Figure 3, and 
one can visualize small Minkowski coordinate frames situated at every point 
along the curve AJ (see K$_{2}$ in Fig. 3) with integration taking place 
along a series of spatial axes (such as X$_{2}$ in Fig. 3). By doing this 
one arrives at a length for AJ, the presumed circumference of a disk of 
radius $r$, of
\begin{equation}
\label{eq1}
{\rm AJ}=\frac{2\pi r}{\sqrt {1-\omega ^2r^2/c^2} }>\,\,{\rm 
Circumference}\,\,{\rm in}\,\,{\rm lab},
\end{equation}
and thus, the disk surface is concluded to be non-Euclidean (Riemann 
curved.)

But consider that since point p moves along a timelike path as seen from K 
(see dotted line in Fig. 3), a time difference between events A at 0$^{o}$ 
and J at 360$^{o}$ must therefore exist as measured by a clock attached to 
point p. To continue from J to A requires a jump in time, and thus, the 
traditional analysis approach leads to a discontinuity in time (or 
alternatively, multi-valued time), a seemingly impossible physical 
situation. Further, as noted by Weber\cite{Weber:1997}, this means that if 
light rays are sent 360$^{o}$ around the rim to synchronize the clock at J 
with that at A, then the two clocks (which are really one and the same 
clock) are not in synchronization. That is, each clock on the disk is out of 
synchronization with itself.

Still further, according to the traditional analysis, time all along the 
path AJ is fixed. Thus, by that analysis, which depends on the locality 
hypothesis and integration of values (time in this case) from local frame to 
local frame, A and J must be simultaneous. But they are not.

\subsubsection{Traditional Approach to Resolve the ``Time Gap''}
In recent years, this problem has been treated as if this ``time gap'' were 
a mathematical artifact, and approaches labeled 
``desynchronization''\cite{Ref:2}$^{,}$\cite{Rizzi:1999}$^{,}$\cite{Rizzi:1} 
and ``discontinuity in 
synchronization''\cite{Cranor:2000}$^{,}$\cite{Anandan:1981} have been 
proposed that entail multiple clocks at a given event. These approaches seem 
motivated by the gauge theory of synchronization philosophy that time 
settings on clocks are inherently arbitrary.

Furthermore, the time gap is often said to be identical in nature to 
traveling at constant radius in a polar coordinate system. The $\phi $ value 
is discontinuous at 360$^{o}$. Similar logic applies for time, with the 
International Date Line for the time zone settings on the earth. If one 
starts at that line and proceeds 360$^{o}$ around the earth, one returns to 
find one must jump a day in order to re-establish one's clock/calendar 
correctly.

\subsubsection{Arguments for Physical Interpretation of the ``Time Gap''}
The gauge synchronization philosophy champions innumerable, equally valid, 
synchronization schemes. Yet, within any one of those schemes, time is 
single valued and continuous, and clocks are all in synchronization with 
themselves. For a given synchronization method, each event within a given 
frame has a single time associated with it.

In the desynchronization approaches, a given event in a given rotating 
frame, for a given synchronization method, can have any number of possible 
times on it. For example, the clock at point p in Fig. 2 has one time on it 
at event A. If one Einstein synchronizes the clock at 360$^{o}$ (i.e., the 
same clock at p) with ccw light rays, one gets another time setting. Thus, 
one has a choice of which of two times one prefers for any given event at 
point p. If, on the other hand, one synchronizes the clock at 360$^{o}$ via 
cw light rays, one gets yet another setting, and a third possible time to 
choose for any given event. Consider yet another path in which the light ray 
goes radially inward 1 meter, then 360$^{o}$ around, then radially outward 1 
meter. One then gets yet another setting for the clock at p. Since there are 
an infinite number of possible paths by which one could synchronize the 
clock at p, there are an infinite number of possible times for each event at 
p. (This does not happen in translation. Any possible path for the light 
rays results in the same unique setting, for a given synchronization scheme, 
on each clock in the frame.)

This plethora of possible settings for the same clock results from insisting 
on ``desynchronization'' of clocks in order to keep the speed of light 
locally $c$ everywhere. And thus, one is in the position of choosing whichever 
value for time one needs in a given experiment in order to get the answer 
one insists one must have (i.e., invariant, isotropic local light speed.) 
One can only then ask if this is really physics or not. Can an infinite 
number of possible readings on a single clock at a single event for a single 
method of synchronization be anything other than meaningless?

The polar coordinate analogy, I believe, confuses physical 
discontinuity with coordinate discontinuity. In 2D, place a green X at 
0$^{o}$, travel 360$^{o}$ at constant radius, and then place a red X. The 
red and green marks coincide in space. There is no discontinuity in space 
between them, although there is a discontinuity in the coordinate $\phi $.

Flash a green light on the equator at the International Dateline, then trace 
a path once around the equator along which no time passes. If you flash a 
red light at the end of that path, the red and green flashes are coincident 
in space and time. There is no physical discontinuity, although your time 
zone clocks show a coordinate discontinuity.

In Fig. 3, flash a green light at event A. Travel 360$^{o}$ on the disk 
along the space-time path AJ (along which no time passes according to the 
traditional analysis), then flash a red light at event J. The two lights are 
not coincident. There is real world space-time gap between them, and they 
exist at different points in 4D. The discontinuity is physical, not merely 
coordinate.

Peres was aware of this time discontinuity, calling it a \textit{``heavy price which we are paying to make the [circumferential] velocity of light ... equal to c''}\cite{Peres:1}. 
Dieks\cite{Dieks:1991} noted that though arbitrary in certain senses, time 
in relativity must be \textit{``directly linked to undoubtedly real physical processes''}. This author agrees.

\subsubsection{The Only Physically Possible Synchronization/Simultaneity}
\label{subsubsec:mylabel2}
There are potential choices for synchronization/simultaneity in the rotating 
frame other than Einstein's. The traditional one with local Einstein 
synchronization around the disk rim is based on the Lorentz transformation 
from the lab to the local co-moving inertial frame, i.e.
\begin{equation}
\label{eq2}
cdT_i =\frac{1}{\sqrt {1-v^2/c^2} }(cdT-\frac{v}{c}dX)=\gamma 
(cdT-\frac{v}{c}dX)
\end{equation}
where $v=\omega r$, \textit{dT} is the time interval in the lab, \textit{dX} is the space interval 
in the lab along the disk rim, \textit{dT}$_{i}$ is the time interval in the local 
co-moving inertial frame, which we presume, by the locality hypothesis, 
equals the time interval on the disk. We could just as well have 
chosen\cite{See:2}
\begin{equation}
\label{eq3}
cdT_i =\gamma (cdT-\kappa dX)
\end{equation}
where $\kappa $ could have any value other than $v/c$.

However, for any $\kappa \ne 0$, we would again have a time discontinuity, 
and all the issues with multiple event times and clocks being out of 
synchronization with themselves of the prior section.

I suggest that, prior to all else, any theory of rotation must be compliant 
with the physical world constraint that time be continuous and single valued 
(within a given frame, and for a given synchronization scheme.) That is only 
possible for a synchronization/simultaneity scheme where $\kappa =0$. For 
this scheme, events in the lab that are simultaneous (i.e., have \textit{dT} = 0 
between them) are also simultaneous in the rotating frame (have \textit{dt} = 0).

\textbf{Postulate:} Any synchronization/simultaneity scheme for the rotating 
frame for which that frame and the lab do not share common simultaneity 
results in a physical time discontinuity and is thus unacceptable on 
physical grounds.

The traditional approach to rotation is at odds with this postulate.

\subsection{Experiment and the Traditional Approach}
In Part 3, virtually all of the experiments that have been carried out to 
verify SRT are reviewed. One of these, the Michelson-Morley type experiment 
performed by Brillet and Hall\cite{Brillet:1979}, found a persistent, 
anomalous, non-null signal at the 10$^{-13}$ level, which is not predicted 
by SRT. The approach to relativistic rotation of Part 2, which is based on 
the above postulate, predicts this signal, and otherwise, makes the same 
predictions as the traditional approach for the remaining tests.

Furthermore, as a result of studies on the Global Positioning System (GPS) 
data for the rotating earth, recognized world leading GPS expert Neil Ashby 
states

\textit{``Now consider a process in which observers in the rotating frame attempt to use Einstein synchronization..... Simple minded use of Einstein synchronization in the rotating frame ... thus leads to a significant error''.}\cite{Ashby:1997}

He also recently noted in \textit{Physics Today},

\textit{`` .. the principle of the constancy of c [the speed of light] cannot be applied in a rotating reference frame ..'' }\cite{Ashby:2002}.

\subsection{Summary of Part 1}
\label{subsec:summary}
Thought experiments, actual experiments, and the physical nature of the 
space-time continuum appear discordant with the traditional approach to 
relativistic rotation.

\section{Resolution of the Conundrums: Differential Geometry and 
Non-time-orthogonality}
\textit{``.. a good part of science is distinguishing between useful crazy ideas and those that are just plain nutty.''}

Princeton University Press advertisement for the book ``Nine Crazy Ideas in 
Science''

\subsection{Introduction}
Part 2 poses an alternative approach to relativistic rotation that resolves 
the inconsistencies, and as will be seen in Part 3, appears to have better 
agreement with experiment than the traditional approach. There are two 
fundamental steps to the alternative approach.

\begin{enumerate}
\item Postulate that, in accord with physical world logic as presented in Part 1, simultaneity/synchronization in the rotating frame can only be such that time in that frame is continuous and single valued.
\item Apply differential geometry and note resulting predictions.
\end{enumerate}

Before beginning the analysis, relevant background material from 
differential geometry is presented in Section \ref{subsec:physical}.

\subsection{Physical vs. Coordinate Components}
\label{subsec:physical}
If one has coordinate components, found from generalized coordinate tensor 
analysis, for some quantity, such as stress or velocity, one needs to be 
able to translate those into the values measured in experiment. For some 
inexplicable reason, the method for doing this is not typically taught in 
general relativity (GR) texts/classes, so it is reviewed here. (Note that 
often in GR, one seeks invariants like $d\tau $, \textit{ds}, etc., which are the same 
in any coordinate system, and in such cases, this issue does not arise. The 
issue does arise with vectors/tensors, whose coordinate components vary from 
coordinate system to coordinate system.)

The measured value for a given vector component, unlike the coordinate 
component, is unique within a given reference frame. In differential 
geometry (tensor analysis), that measured value is called the ``physical 
component''.

Many tensor analysis texts show how to find physical components from 
coordinate components\cite{The:1972}. A number of continuum mechanics texts 
do as well\cite{Malvern:1988}. The only GR text known to the present author 
that mentions physical components is Misner, Thorne, and 
Wheeler\cite{See:3}. Those authors use the procedure to be described below, 
but do not derive it\cite{Physical:1}. The present author has written an 
introductory article on this, oriented for students, that may be found at 
the Los Alamos web site\cite{Klauber:1}. The following is excerpted in part 
from that article.

The displacement vector $d$\textbf{x }between two points in a 2D Cartesian 
coordinate system is
\begin{equation}
\label{eq4}
d{\rm {\bf x}}\,=dX^1{\rm {\bf \hat {e}}}_1 +dX^2{\rm {\bf \hat {e}}}_2 
\end{equation}
where the ${\rm {\bf \hat {e}}}_i $ are unit basis vectors and \textit{dX}$^{i}$ are 
physical components (i.e., the values one would measure with meter sticks). 
For the same vector $d$\textbf{x} expressed in a different, generalized, 
coordinate system we have different coordinate components \textit{dx}$^{i }\ne $ 
\textit{dX}$^{ i }$ (\textit{dx}$^{i}$ do not represent values measured with meter sticks), but a 
similar expression
\begin{equation}
\label{eq5}
d{\rm {\bf x}}\,=dx^1{\rm {\bf e}}_1 +dx^2{\rm {\bf e}}_2 ,
\end{equation}
where the generalized basis vectors \textbf{e}$_{i}$ point in the same 
directions as the corresponding unit basis vectors ${\rm {\bf \hat {e}}}_i 
$, but are not equal to them. Hence, for ${\rm {\bf \hat {e}}}_i $, we have
\begin{equation}
\label{eq6}
{\rm {\bf \hat {e}}}_i =\frac{{\rm {\bf e}}_i }{\vert {\rm {\bf e}}_i \vert 
}=\frac{{\rm {\bf e}}_i }{\sqrt {{\rm {\bf e}}_{\underline{i}} \cdot {\rm 
{\bf e}}_{\underline{i}} } }=\frac{{\rm {\bf e}}_i }{\sqrt 
{g_{\underline{i}\underline{i}} } }
\end{equation}
where underlining implies no summation.

Substituting (\ref{eq6}) into (\ref{eq4}) and equating with (\ref{eq5}), one obtains
\begin{equation}
\label{eq7}
dX^1=\sqrt {g_{11} } dx^1\quad \quad \quad dX^2=\sqrt {g_{22} } dx^2,
\end{equation}
which is the relationship between displacement physical (measured with 
instruments) and coordinate (mathematical value only) components.

Consider a more general case of an arbitrary vector \textbf{v} 
\begin{equation}
\label{eq8}
{\rm {\bf v}}=v^1{\rm {\bf e}}_1 +v^2{\rm {\bf e}}_2 =v^{\hat {1}}{\rm {\bf 
\hat {e}}}_1 +v^{\hat {2}}{\rm {\bf \hat {e}}}_2 
\end{equation}
where, \textbf{e}$_{1}$ and \textbf{e}$_{2}$ here do not, in general, have 
to be orthogonal, \textbf{e}$_{i}$ and ${\rm {\bf \hat {e}}}_i $ point in 
the same direction for each index $i$, and carets over component indices 
indicate physical components. Substituting (\ref{eq6}) into (\ref{eq8}), one readily obtains
\begin{equation}
\label{eq9}
v^{\hat {i}}=\sqrt {g_{\underline{i}\underline{i}} } v^i,
\end{equation}
which we have shown here to be \textit{valid in both orthogonal and non-orthogonal systems}.

As a further aid to those readers familiar with anholonomic coordinates 
(which are associated with non-coordinate basis vectors superimposed on a 
generalized coordinate grid), physical components are special 
case anholonomic components for which the non-coordinate basis vectors have unit 
length.

It is important to recognize that anholonomic components do not transform as true 
vector components. So one can not simply use physical components in tensor 
analysis as if they were. Typically, one starts with physical components as 
input to a problem. These are converted to coordinate components, and the 
appropriate tensor analysis carried out to get an answer in terms of 
coordinate components. One then converts these coordinate components into 
physical components as a last step, in order to compare with values measured 
with instruments in the real world.

As a basis vector is derived from infinitesimals (derivative at a point), 
one sees (\ref{eq9}) is valid locally in curved, as well as flat, spaces, and can be 
extrapolated to 4D general relativistic applications. So, very generally, 
for a 4D vector $v^{\mu }$ and a metric signature (-,+,+,+)
\begin{equation}
\label{eq10}
v^{\hat {i}}=\sqrt {g_{\underline{i}\underline{i}} } v^i\quad \quad v^{\hat 
{0}}=\sqrt {-g_{00} } v^0,
\end{equation}
where Roman sub and superscripts refer solely to spatial components (i.e. 
$i$ = 1,2,3.)

\subsection{Alternative Analysis Approach}
\label{subsec:alternative}
We begin with the simultaneity postulate of Section 
\ref{subsubsec:mylabel2}, repeated below for convenience.

\textbf{Postulate:} Any synchronization/simultaneity scheme for the rotating 
frame, for which that frame and the lab do not share common simultaneity, 
results in a physical time discontinuity and is thus unacceptable on 
physical grounds.

\subsubsection{Disk Transformation and Metric}
As will be discussed, the global transformation from the lab to the rotating 
frame apparently first used by Langevin\cite{Langevin:1937} to find a 
suitable metric for the rotating frame incorporates the above postulate. 
This transformation is used in the following analysis, which parallels that 
of Klauber\cite{Klauber:1998}. The correctness of the transformation can be 
judged by whether the predictions made by using it match experiment.

This transformation, between the non-rotating (lab, upper case symbols) 
frame to a rotating (lower case) frame, is
\begin{equation}
\label{eq11}
\begin{array}{l}
 cT=ct \\ 
 R=r \\ 
 \Phi =\phi +\omega t \\ 
 Z=z\,. \\ 
 \end{array}
\end{equation}
$\omega $ is the angular velocity of the rotating frame as seen from the lab, 
and cylindrical spatial coordinates are used. The coordinate time $t$ for the 
rotating system equals the proper time of a standard clock located in the 
lab. Substituting the differential form of (\ref{eq11}) into the line element in the 
lab frame
\begin{equation}
\label{eq12}
ds^2 = - c^2dT^2+ dR^2 + R^2d\Phi ^2{\rm 
}+ dZ^2
\end{equation}
results in the line element for the rotating frame
\begin{equation}
\label{eq13}
ds^2\;\;=\;-c^2(1-\textstyle{{r^2\omega ^2} \over 
{c^2}})dt^2\;+\;dr^2\;+\;r^2d\phi ^2\;+\;2r^2\omega d\phi 
dt\;+\;dz^2=\;g_{\alpha \beta } dx^\alpha dx^\beta .
\end{equation}
Note that the metric in (\ref{eq13}) is not diagonal, since $g_{\phi t} \ne 0$, and 
this implies that time is not orthogonal to space (i.e., a 
non-time-orthogonal, or NTO, frame.)

\subsubsection{Time on the Disk}
\label{subsubsec:mylabel3}
Time on a standard clock at a fixed 3D location on the rotating disk, found 
by taking \textit{ds}$^{2}= -c^{2}d\tau $ and \textit{dr = d}$\phi $ = \textit{dz = }0 in (\ref{eq13}), is
\begin{equation}
\label{eq14}
d\tau =\sqrt {1-r^2\omega ^2/c^2} dt=\sqrt {-g_{tt} } dt,
\end{equation}
This varies with radial position $r$. At the axis of rotation (where $r$ = 0), the 
standard clock agrees with the clock in the lab. At other locations, 
standard clock time is diluted by the Lorentz factor, as in traditional SRT. 
The coordinate time everywhere on the disk is $t$, and that equals the time $T$ in 
the lab.

The time difference between two events at two locations (each having its own 
clock) on the disk, in coordinate components, is \textit{dt}. The corresponding 
physical time difference is
\begin{equation}
\label{eq15}
dt_{phys} =d\hat {t}=\sqrt {-g_{tt} } dt=\sqrt {1-r^2\omega ^2/c^2} dt.
\end{equation}
If the two locations happen to be one and the same location, one obviously 
gets (\ref{eq14}).

Note that two events seen as simultaneous in the lab have \textit{dT} = 0 between them. 
From the first line of (\ref{eq11}) and (\ref{eq15}), the same two events must also have 
$dt_{phys} =0$, and thus they are also seen as simultaneous on the disk. 
This statement is true for all standard (physical) clocks on the disk. 
Though the standard clocks at different radii thereon run at different 
rates, and thus can not be synchronized, they \textit{can} share common simultaneity. 
The lab shares this common simultaneity with all of the disk clocks, and 
thus our postulate above holds for transformation (\ref{eq11}), and the resulting 
(NTO) metric of (\ref{eq13}).

Note further, that the simultaneity chosen here is equivalent in the 
physical world to what is sometimes called ``flash from center'' 
simultaneity (or synchronization if one is confined to clocks at fixed 
radius). In that scheme, one imagines a flash of light on the axis of 
rotation whose wave front propagates outwardly in all radial directions. 
Events when the wave front impacts individual points along a given 
circumference are considered simultaneous.

It is significant that the ``flash from center'' synchronization is the same 
as that proposed (via other logic) near the end of Section 
\ref{subsubsec:thought}.

\subsubsection{Local Speed of Light on the Disk}
\label{subsubsec:local}
For light \textit{ds}$^{2}$ = 0. Inserting this into (\ref{eq13}), taking \textit{dr=dz=}0, and using the 
quadratic equation formula, one obtains a local coordinate velocity 
(generalized coordinate spatial grid units per coordinate time unit) in the 
circumferential direction
\begin{equation}
\label{eq16}
v_{light,coord,circum} =\frac{d\phi }{dt}=-\omega \pm \frac{c}{r},
\end{equation}
where the sign before the last term depends on the circumferential direction 
(cw or ccw) of travel of the light ray. The local physical velocity (the 
value one would measure in experiment using standard meter sticks and clocks 
in units of meters per second) is found from this to be
\begin{equation}
\label{eq17}
v_{light,phys,circum} \;\;=\;\;\frac{\sqrt {g_{\phi \phi } } d\phi }{\sqrt 
{-g_{tt} } dt}\;\;=\;\;\frac{-\;r\omega \;\pm \;c}{\sqrt 
{1-\textstyle{{\omega ^2r^2} \over {c^2}}} }=\;\;\frac{-\;v\;\pm \;c}{\sqrt 
{1-\textstyle{{v^2} \over {c^2}}} }.
\end{equation}
Note that for this approach, the local physical speed of light in rotating 
frames is not invariant or isotropic, and that this lack of 
invariance/isotropy depends on $\omega $, the angular velocity seen from the 
lab. Note particularly that this result is a direct consequence of the NTO 
nature of the metric in (\ref{eq13}). If $\omega $=0, local physical (measured) 
light speed is isotropic and invariant, the metric is diagonal, and time is 
orthogonal to space.

I thus call this alternative analysis method, the NTO approach to 
relativistic rotation.

\subsection{Implications of NTO Approach}
\label{subsec:implications}
\subsubsection{Hypothesis of Locality}
Local physical light speed in the rotating frame, according to the NTO 
approach, is not equal to $c$. Yet, in a local, co-moving, Lorentz 
frame, which via the hypothesis of locality is equivalent locally to the 
non-inertial frame, the physical speed of light is always $c$. Thus, a local 
co-moving Lorentz frame is \textit{not} equivalent locally to the rotating frame, and 
the hypothesis of locality is not valid for such frames. One does \textit{not} measure 
the same values for velocity, and hence by implication for time and space, 
in the two frames.

This is a direct result of the simultaneity postulate, required to keep time 
in the rotating frame continuous and single-valued. That requirement results 
in a rotating frame that can only be NTO, i.e., it can only have a metric 
with off diagonal terms in the metric.

I submit that the hypothesis of locality remains true only for those 
non-inertial frames in which it is possible for the metric to have all null 
off-diagonal space-time components. This set of frames comprises the vast 
majority of problems encountered in GR. Rotation is a critical exception.

\subsubsection{Absolute Nature of Simultaneity in Rotation}
In NTO analysis, simultaneity/synchronization on the rotating disk, unlike 
that in the gauge theory of synchronization, is unique (absolute.) The gauge 
theory validity, it is submitted, is restricted to translating frames and 
does not apply to rotation. This is not unlike other differences between 
rotation and translation. Velocity in rotation, for example, has an absolute 
quality, whereas in translation it does not. There is a preferred frame in 
rotation, upon which everyone agrees (the frame with no Coriolis effect, for 
example); in translation, there is no such frame.

\subsubsection{Lorentz Contraction Revisited}
To determine Lorentz contraction of meter sticks, we merely need to compare 
physical length in the circumferential direction in both the lab and 
rotating frames, i.e., look at the physical component for \textit{d$\Phi $} and \textit{d$\phi $}. This is 
equivalent to finding the proper length when \textit{dT =} 0 in the first frame (lab 
here), and\textit{ dt} = 0 in the second frame (disk here), which is what one does in 
SRT.

The distance between two points along the circumference in the lab in meter 
sticks is
\begin{equation}
\label{eq18}
d\Phi _{phys} =d\hat {\Phi }=\sqrt {g_{\Phi \Phi } } d\Phi =Rd\Phi =R(\Phi 
_2 -\Phi _1 ),
\end{equation}
which is not surprising, and which (for \textit{dR = dT = dZ =} 0) equals \textit{ds}. (\ref{eq18}) represents the 
number of meter sticks between points 1 and 2 in the lab. Now, consider two 
3D points on the disk located instantaneously at the same place as points 1 
and 2 in (\ref{eq18}). We ask, how many meter sticks span that distance as measured 
on the disk? That distance between points 1 and 2 in meter sticks is
\begin{equation}
\label{eq19}
d\phi _{phys} =d\hat {\phi }=\sqrt {g_{\phi \phi } } d\phi =rd\phi =r(\phi 
_2 -\phi _1 ).
\end{equation}
According to (\ref{eq11}), $\phi _1 =\Phi _1 -\omega t_1 $ and $\phi _2 =\Phi _2 
-\omega t_2 $. Since, $r = R$ and $dt=t_2 -t_1 =0$, (\ref{eq18}) and (\ref{eq19}) are equal. The 
disk observer sees the same number of meter sticks between two points as the 
lab observer does between those points, and hence, there is no Lorentz 
contraction.

Note that we would need a metric component $g_{\phi \phi } \ne r^2$ in the 
rotating frame to have Lorentz contraction. The postulate of 
simultaneity/time continuity leads to the metric of (\ref{eq13}), which is NTO, and 
which has $g_{\phi \phi } =r^2$.

The Lorentz contraction issue is treated more extensively, and with 
graphical illustrations, in Klauber\cite{Klauber:1998}. The limit case for 
NTO analysis is also discussed therein though it is treated at great length 
in Klauber\cite{Klauber:2}, and found to be free of inconsistencies.

\subsubsection{Sagnac and Thought Experiments}
\label{subsubsec:sagnac}
A complete and general derivation of the Sagnac result from the rotating 
frame using NTO analysis can be found in Klauber\cite{Klauber:3}. Shown 
below is the simpler derivation for a circumferential light path whose 
center is the axis of rotation. Note that different speeds for light in the 
cw and ccw directions is inherent in the NTO approach, and thus that 
approach is completely consonant with the thought experiment of Part 1, 
Section \ref{subsubsec:thought}.

The difference in time measured on a ccw rotating disk between two pulses of 
light traveling opposite directions along a circumferential arc of length 
\textit{dl} is
\begin{equation}
\label{eq20}
dt_{phys} =\frac{dl}{v_- }-\frac{dl}{v_+ },
\end{equation}
where $v_{+}$ is the speed for the cw light ray and $v_{-}$ is the speed for 
the ccw ray. Using (\ref{eq17}) this becomes
\begin{equation}
\label{eq21}
dt_{phys} =\frac{dl\sqrt {1-v^2/c^2} }{c-v}-\frac{dl\sqrt {1-v^2/c^2} 
}{c+v}=\frac{v}{c^2}\frac{2dl}{\sqrt {1-v^2/c^2} }.
\end{equation}
By integrating the RHS of (\ref{eq21}) from 0 to 2$\pi r$ (recall there is no Lorentz 
contraction), the LHS becomes the time difference on the clock fixed at the 
emission/reception point,
\begin{equation}
\label{eq22}
\Delta t_{phys} =\frac{\omega r}{c^2}\frac{2(2\pi r)}{\sqrt {1-\omega 
^2r^2/c^2} }=\frac{4\omega A}{c^2\sqrt {1-\omega ^2r^2/c^2} },
\end{equation}
which agrees with the derivation from the lab frame of the Appendix.

\subsubsection{Brillet and Hall}
\label{subsubsec:brillet}
The Brillet and Hall\cite{Brillet:1979} experiment is described in Part 3. 
It remains to this day the only test of sufficient accuracy to detect any 
non-null Michelson-Morley (MM) effect due to the surface speed of the earth 
rotating about its axis. Brillet and Hall found null signals for the solar 
and galactic orbit speeds. However, they noted a persistent non-null signal 
at 2 X 10$^{-13}$, which had fixed phase in the lab frame.

This signal is not predicted by traditional SRT, which insists on local 
Lorentz invariance for light speed, and was thus simply deemed ``spurious'' 
without further explanation. However, this signal is predicted by NTO 
analysis due to the earth surface speed. (See Klauber\cite{Klauber:4}.)

\subsubsection{Gravitational Orbit vs. True Rotation}
One could ask why any test should get a null signal for the solar and 
galactic orbital velocities, but a non-null signal for the earth surface 
speed from its own rotation.

The answer is that a body in gravitational orbit is in free fall, and is 
therefore Lorentzian. No centrifugal ``force'' is felt, and no Foucault 
pendulum moves, as a result of revolution in orbit. There is no experimental 
means by which one could determine (without looking outside at the stars) 
one's rate of revolution in orbit. Hence, you can not determine any absolute 
circumferential speed, and the second postulate of relativity holds. Related 
logic\cite{Klauber:5} leads to the conclusion that the speed of light on 
such a body is invariant and equal to $c$ as well.

Thus, the usual form of relativity should hold for gravitational orbits and 
one should expect a null Michelson-Morley result for orbital speeds, which 
is just what is measured. However, one \textit{can} use instruments to determine the 
speed of the earth's surface about its axis, and therefore we should expect 
that relativity theory will not hold in precisely the same form for that 
case. It is submitted that the Brillet and Hall result justifies that 
expectation.

This subject is treated in depth in Klauber\cite{Klauber:5}.

\subsubsection{NTO vs. Selleri Transformations}
\label{subsubsec:mylabel4}
In treating rotation, Selleri\cite{Selleri:1998} uses the same simultaneity 
as the lab (though he advocates an ``absolute'' simultaneity that pervades 
translation as well.) He finds anisotropic one-way light speed on a rotating 
disk as
\begin{equation}
\label{eq23}
v_{light,phys,circum,Selleri} =\frac{-\omega r\pm c}{1-\omega ^2r^2/c^2}
\end{equation}
for the cw and ccw speeds of light along the circumference. Comparing this 
with the NTO relation of (\ref{eq17}), one finds the two differ by a factor of 
$1/\sqrt {1-\omega ^2r^2/c^2} $.

Selleri shows that his relation (\ref{eq23}) results in a circular round trip speed 
for light (one way around the rim) that agrees with (the first order) Sagnac 
experimental results. However, for a back and forth round trip for light 
along the same path, he shows his relation results in a round trip speed of 
precisely $c$. Thus, he predicts a null result for any Michelson-Morley type 
experiment.

NTO analysis on the other hand, due to the Lorentz factor difference from 
(\ref{eq23}), predicts a back and forth round trip speed for light as not equal to 
$c$. Therefore, it predicts a non-null result for MM experiments (which are 
sensitive enough to detect effects from the earth surface speed due to its 
own rotation.)

This difference can be attributed to the lack, in the NTO approach, of 
circumferential Lorentz contraction, as opposed to the inclusion of such 
contraction in the Selleri approach. Given Lorentz contraction, light rays 
will travel a greater number of meter sticks, and thus speed will be 
increased by the magnitude of that contraction. This is the difference 
between (\ref{eq23}) and (\ref{eq17}).

\subsubsection{Co-moving vs. Disk-fixed Observers}
\label{subsubsec:mylabel5}
It should be clearly noted that in the NTO approach, the rotating disk fixed 
observer and the local co-moving Lorentz observer are not equivalent. They 
do not, for example, see the lab meter sticks as having the same length. 
This is in accord with earlier statements regarding the invalidity of the 
hypothesis of locality for rotating frames.

From another perspective, it could be claimed that the two observers are not 
truly co-moving, as the disk observer at \textit{r} is rotating (at $\omega )$, whereas 
the local Lorentz observer is not.

\subsection{Summary of Part 2}
By adopting

1) the postulate that time in a rotating frame must be continuous and single 
valued (each clock must be in synchronization with itself), and

2) the specific transformation of form (\ref{eq11}) that incorporates that 
postulate,

\noindent
one can develop an NTO theory for rotation that resolves all conundrums of 
Part 1, and in which the physical speed of light is constrained to be 
locally anisotropic. One finds agreement with experiment, including the 
Brillet and Hall test result, which is not predicted by the traditional 
approach to relativistic rotation.

One also finds the hypothesis of locality can only be true for non-inertial 
frames in which the metric can be expressed in diagonal form and still 
maintain continuity in time. In rotation, this is not true, and the local 
co-moving observer does not see events (in particular, meter stick lengths) 
in the same way as the disk-fixed observer.

\section{Experiment and Non-time-orthogonal Analysis}
\subsection{Introduction}
Part 3 reviews the experiments that have been performed to test special 
relativity, and implicitly therein, the hypothesis of locality and the 
traditional approach to relativistic rotation. Results of these experiments 
are examined in order to compare the predictive capacity of the NTO and 
traditional analysis approaches.

\subsection{The Experiments}
Table 1 is an extensive list of experiments performed since 1887 capable of 
evaluating at least one aspect of SRT. Particular experiments are referred 
to herein via the symbol in the first column. A terse description of each is 
given in column two, with the year and author citations in column three. Note the acronym SRT implies 
both special relativity theory and the traditional approach to rotation. 
Column four briefly summarizes how the NTO effect in a given experiment 
compares with the traditional approach effect. The 
last two columns compare the predictions of NTO and the traditional approach 
(Trad) for the given experiment. For a summary of JPL, M\"{o}ssbauer, TPA, 
and GPA, see Will\cite{Will:1992}. For a summary of Hughes-Drever, BH, NBS, 
UWash, and M\"{o}ssbauer see Haugan and Will\cite{Haugan:1987}.

Three experiments known to the author are not included in the table because 
their results were contrary to both SRT and NTO theories. What these results 
mean is subject to debate, though most physicists who are aware of them 
believe they must be in error. The earliest of these was by 
Miller\cite{Miller:1933}, a highly respected colleague of Michelson. He 
repeated the Michelson-Morley test four times over many years, with various 
equipment in various places, and much of the work was done jointly with 
Morley. The other experiments reporting results contrary to SRT were by 
Silvertooth\cite{Silvertooth:1992} and Marinov\cite{Marinov:1985}. In any 
event, these experiments do not discern between the Trad and NTO approaches, 
and are referenced here for completeness.

\subsection{The Comparisons}
Both the traditional and NTO approaches predict time dilation, and 
experiments measuring this, such as PartAcc (see table), would, for the most 
part, provide no capability of differentiating between approaches. Also, 
Doppler shift effects tend to be the same in NTO and Trad, though, for 
certitude, each experiment comprising Doppler measurement needs to be 
evaluated on its own. 

Tests of the speed of light itself, such as MM, should be more directly 
indicative. These must, however, have i) sufficient accuracy to detect any 
effect from the relatively low earth surface speed about its axis, and ii) 
apparatus that turns with respect to the earth surface. The MM, Post MM, and 
Joos tests, for example, lacked the first of these. The JPL and CORE 
experiments lacked the second. The LFV test did not meet either criterion.

For some tests, there is uncertainty. For example, in the ODM experiment, 
rotation of the apparatus yielded a persistent $\sim $1.5 km/sec variation, 
which was attributed to the earth's magnetic field. This would, however, 
mask any NTO effect (at $\sim $0.35 km/sec), and yield uncertainty as to 
whether Trad predicts the result or not. In the Hughes-Drever test, Doppler 
shifts are measured and NTO usually predicts the same shift as Trad. 
Extensive analysis would be required, however, to be certain.

The most interesting of the tests is BH (Brillet and Hall), as this is the 
only one for which NTO and Trad differ with certainty with regard to results. 
BH used a Fabry-Perot interferometer that rotated with respect to the lab. A 
fraction of the light ray incident on the interferometer emerged directly 
from the far end. Another portion of the ray was reflected at the far end 
and forced to travel round trip, rear to front to rear, before emerging. The 
different portions interfered to form a fringe pattern. If the round trip 
speed of light were anisotropic, the time for it to travel back and forth 
inside the interferometer would vary with orientation of the apparatus. 
This, in turn, would cause the fringe pattern, and thus, the signal BH 
monitored, to vary. In Newtonian theory, this variation, peak-to-peak and to 
second order, is $\textstyle{1 \over 2}v^2/c^2$, where $v$ is the maximum 
change from $c$ of the speed of light. As shown by Klauber\cite{Klauber:4}, 
the NTO effect on light transit time is quantitatively the same (though 
subtle calculational differences exist from the Newtonian analysis.)

The speed of the earth surface about its axis at the location of the BH test 
is .355 km/sec. For this, the amplitude of the variation via NTO theory 
should be
\begin{equation}
\label{eq24}
\frac{1}{4}\frac{v^2}{c^2}=\frac{1}{4}\left( {\frac{.355}{3\times 10^5}} 
\right)^2=3.5\times 10^{-13}
\end{equation}
at twice the apparatus rotation rate. The BH test found a ``persistent'' 
$\sim 1.9\times 10^{-13}$ signal at that rate and with fixed phase relative 
to the earth surface. They deemed this signal ``spurious'', because it 
seemed inexplicable. The character of the BH signal and its proximity in 
value to (\ref{eq24}) should, of themselves, be intriguing. However, there is a 
secondary effect of light speed anisotropy on the BH signal.

The path of travel is altered slightly when the light ray direction is 
transverse to the principle direction of anisotropy. In a heuristic sense, 
the ray seems to be pushed ``sideways''. In the BH experiment, this would 
result in a shifting of the fringe pattern, and a concomitant change in the 
measured signal. Klauber\cite{Klauber:4} calculated this effect and 
found it dependent on certain dimensions of the apparatus, which are not 
known. However, by using values for these dimensions estimated from the 
figure of the apparatus shown in the BH report, he found an expected net 
signal from all effects of $\sim 2\times 10^{-13}$.

\subsection{Comparison to Selleri}
\label{subsec:mylabel2}
As noted in Section \ref{subsubsec:mylabel4}, the Selleri theory, like 
the NTO approach, is based on what this author considers a physically 
defensible simultaneity scheme. The Selleri theory, on the other hand, 
predicts a null signal for the BH experiment.

It would be interesting to compare predictions of the Selleri theory to 
results of other tests such as M\"{o}ssbauer.

\subsection{GPS and Sagnac}
I do not profess expertise in the GPS system, though I have noted 
earlier the remarks by Ashby, who does have extensive expertise. Those 
remarks appear consonant with NTO analysis and its requisite non-Einstein 
synchronization and local light speed anisotropy.

Furthermore, in the context of the thought experiment of Section 
\ref{subsubsec:thought}, the traditional approach seems incapable of 
deriving the Sagnac effect from within the rotating frame. That is, 
considering the local physical speed of light to be isotropic does not seem 
sufficient to derive different arrival times for the cw and ccw light rays. 
This is not the case for the NTO approach, and in this context, the Sagnac 
experiment may be considered empirical support for it.

\subsection{Future Experiments}
Tobar\cite{Tobar:2002}\cite{Wolf Tobar et al:2002}  (WSMR in Table 1) expects to complete a modified 
version of the Michelson-Morley experiment, accurate to several orders of 
magnitude beyond that of BH, by the end of 2004. He will use a whispering 
spherical mode resonator and rotate it with respect to the lab. Preliminary 
analysis by the present author suggests that the WSMR experiment may be 
capable of detecting an NTO effect on light speed, if it exists, due to the 
surface speed of the earth.

\subsection{Summary of Part 3}
\label{subsec:mylabel3}
Only one non GPS/Sagnac experiment appears capable of distinguishing between 
the traditional and NTO approaches to relativistic rotation, that of Brillet 
and Hall. That test, sensitive to 10$^{-15}$, found a signal at $\sim 
1.9\times 10^{-13}$, which is strikingly close to the signal predicted by 
the NTO approach from the earth surface speed about its axis of rotation, 
and which is not predicted by the traditional approach.

\begin{center}
\textbf{Table 1. History of SRT Experiments}
\end{center}

\begin{longtable}[htbp]
{|p{39pt}|p{108pt}|p{87pt}|p{87pt}|p{28pt}|p{31pt}|l|}
\hline
\endhead
\hline
\endfoot
\begin{center}
\underline {\textbf{Symbol}} \end{center} & 
\begin{center}
\underline {\textbf{Test Description}} \end{center} & 
\begin{center}
\underline {\textbf{Authors (Year)}} \end{center} & 
\begin{center}
\underline {\textbf{NTO Effect}} \end{center} & 
\begin{center}
\underline {\textbf{Trad}} \end{center} & 
\begin{center}
\underline {\textbf{NTO}} \end{center} & 
 \\
\hline
\begin{center}
MM \end{center} & 
Original Michelson-Morley experiment& 
Michelson {\&} Morley\cite{Michelson:1} (1887)& 
Accuracy too low. $\sim $7-10 km/sec.& 
\begin{center}
Y \end{center} & 
\begin{center}
Y \end{center} & 
 \\
\hline
\begin{center}
WW \end{center} & 
Electric field effect of rotating magnetic insulator in magnetic field& 
Wilson and Wilson\cite{Wilson:1913} (1913); Hertzberg et 
al\cite{Hertzberg:2001} (2001)& 
NTO prediction = Trad\cite{Burrows:1997}& 
\begin{center}
Y \end{center} & 
\begin{center}
Y \end{center} & 
 \\
\hline
\begin{center}
Post MM \end{center} & 
Repeats of MM interferometer tests& 
Kennedy (1926); Piccard {\&} Stahel (1926-8); Michelson et al ((1929)& 
Null results: 1 km/sec to 7 km/sec accuracy& 
\begin{center}
Y \end{center} & 
\begin{center}
Y \end{center} & 
 \\
\hline
\begin{center}
Joos \end{center} & 
Version of MM& 
Joos\cite{Joos:1930} (1930)& 
Accuracy too low. $\sim $1.5 km/sec.& 
\begin{center}
Y \end{center} & 
\begin{center}
Y \end{center} & 
 \\
\hline
\begin{center}
KT \end{center} & 
Original experiment on time dilation& 
Kennedy and Thorndike\cite{Kennedy:1932}(1932)& 
Not rotated. Low accuracy $\sim $10 km/sec& 
\begin{center}
Y \end{center} & 
\begin{center}
Y \end{center} & 
 \\
\hline
\begin{center}
Ives-Stilwell \end{center} & 
Doppler frequency time dilation in H canal rays& 
Ives and Stilwell\cite{Ives:1941}(1938, 1941)& 
Accuracy 100X too low for NTO effect\cite{There:1}& 
\begin{center}
Y \end{center} & 
\begin{center}
Y \end{center} & 
 \\
\hline
\begin{center}
PartAcc \end{center} & 
Particle accelerator time dilation on half lives& 
Mid 1900s to present& 
NTO prediction = Trad& 
\begin{center}
Y \end{center} & 
\begin{center}
Y \end{center} & 
 \\
\hline
\begin{center}
ODM \end{center} & 
Two opposite direction NH$_{3}$ maser beams. Ether wind Doppler variation as rotate.& 
Cedarholm et al\cite{Cedarholm:1958} (1958)& 
Rotated 180$^{o}$, $\sim $1.5 km/sec variation. Attributed to earth mag field. & 
\begin{center}
? \end{center} & 
\begin{center}
Y \end{center} & 
 \\
\hline
\begin{center}
Hughes-Drever \end{center} & 
Isotropy of nuclear energy levels. Doppler shift of photons emitted by 2 different atoms.& 
Hughes et al\cite{Hughes:1961} and Drever (1960)& 
Significant analysis needed for NTO prediction.& 
\begin{center}
Y \end{center} & 
\begin{center}
? \end{center} & 
 \\
\hline
\begin{center}
PDM \end{center} & 
Perpendicular direction He-Ne masers. Rotated.& 
Jaseva, Townes et al\cite{Jaseva:1964} (1964)& 
Accuracy too low. Systematic signal as rotated.& 
\begin{center}
? \end{center} & 
\begin{center}
Y \end{center} & 
 \\
\hline
\begin{center}
M\"{o}ss-bauer \end{center} & 
M\"{o}ssbauer rotor. Classical frequency shift different from SRT& 
Champeney et al\cite{Champeney:1963} (1963); Turner and 
Hill\cite{Turner:1964} (1964)& 
NTO predicts same frequency change as Trad& 
\begin{center}
Y \end{center} & 
\begin{center}
Y \end{center} & 
 \\
\hline
\begin{center}
HK \end{center} & 
Time dilation on atomic clocks flown around world& 
Hafele and Keating\cite{Hafele:1972} (1972)& 
NTO prediction = Trad& 
\begin{center}
Y \end{center} & 
\begin{center}
Y \end{center} & 
 \\
\hline
\begin{center}
BH \end{center} & 
Fringe shift in interferometer as rotate& 
Brillet and Hall\cite{Brillet:1979} (1978)&
2$^{nd}$ order effect at 10$^{-13}$. NTO predicts\cite{Klauber:4}& 
\begin{center}
N \end{center} & 
\begin{center}
Y \end{center} & 
 \\
\hline
\begin{center}
GPA \end{center} & 
Gravity probe A. Maser on rocket and maser on ground. Classical Doppler varies from SRT.& 
Vessot and Levine\cite{Vessot:1979} (1979), Vessot et al\cite{Vessot:1980} 
(1980)& 
NTO shift = Trad. Analysis done in non-rotating earth centered frame.& 
\begin{center}
Y \end{center} & 
\begin{center}
Y \end{center} & 
 \\
\hline
\begin{center}
Refract \end{center} & 
Light split in air and glass. 1st order fringe effect in Galilean {\&} Fresnel ether drag theories.& 
Byl et al\cite{Byl:1985} (1985)& 
1$^{st}$ order effect NTO = Trad $\ne $ Galilean or Fresnel drag.\cite{KlauberByl}& 
\begin{center}
Y \end{center} & 
\begin{center}
Y \end{center} & 
 \\
\hline
\begin{center}
NBS \end{center} & 
Isotropy of nuclear energy levels. Doppler shift. Rotation of earth changed orientation.& 
Prestage\cite{Prestage:1985} et al (1985) at National Bureau Standards& 
Apparatus not rotated. NTO effect = Trad& 
\begin{center}
Y \end{center} & 
\begin{center}
Y \end{center} & 
 \\
\hline
\begin{center}
TPA \end{center} & 
2 photon absorption in atomic beam. Doppler shift opposite directions affected by ether wind.& 
Kaivola et al\cite{Kaivola:1985} (1985); Riis et al\cite{Riis:1989} (1988)& 
Apparatus not rotated. Beam aligned N-S\cite{With:1}. NTO effect = Trad& 
\begin{center}
Y \end{center} & 
\begin{center}
Y \end{center} & 
 \\
\hline
\begin{center}
UWash \end{center} & 
Isotropy of nuclear energy levels. Doppler shift. Rotation of earth changed orientation.& 
Lamoreaux et al\cite{Lamoreaux:1986} at Univ Washington (1986)& 
Apparatus not rotated. NTO effect = Trad&

\begin{center}
Y \end{center} & 
\begin{center}
Y \end{center} & 
 \\
\hline
\begin{center}
JPL \end{center} & 
Jet Propulsion Lab. 2 earth fixed masers. Fiberoptic comparison.& 
Krisher et al\cite{Krisher:1990} (1990)& 
Apparatus not rotated. NTO effect = Trad& 
\begin{center}
Y \end{center} & 
\begin{center}
Y \end{center} & 
 \\
\hline
\begin{center}
LFV \end{center} & 
Laser frequency variation as earth rotates: stabilized laser compared to stable cavity locked laser.& 
Hils and Hall\cite{Hils:1990} (1990)& 
Apparatus not rotated, plus accuracy too low for NTO effect.& 
\begin{center}
Y \end{center} & 
\begin{center}
Y \end{center} & 
 \\
\hline
\begin{center}
Sat \end{center} & 
GPS satellite test of SRT.& 
Wolf and Petit\cite{Wolf:1997} (1997)& 
Analysis done in non-rotating earth centered frame. NTO effect = Trad.& 
\begin{center}
Y \end{center} & 
\begin{center}
Y \end{center} & 
 \\
\hline
\begin{center}
CORE \end{center} & 
Cryogenic Optical Resonators measure anisotropy of light speed as earth rotates.& 
Braxmaier et al\cite{Braxmaier:2002} (2002)&
Apparatus not rotated. NTO effect = Trad.& 
\begin{center}
Y \end{center} & 
\begin{center}
Y \end{center} & 
 \\
\hline
\begin{center}
WSMR \end{center} & 
Whispering spherical mode resonator Michelson-Morley experiment.& 
Tobar\cite{Tobar:2002}\cite{Wolf Tobar et al:2002} (2004)& 
Appears capable of discerning between NTO and Trad& 
\begin{center}
TBD \end{center} & 
\begin{center}
TBD \end{center} & 
\label{tab1}
\end{longtable}

\section*{Appendix. Deriving Sagnac Result from the Lab Frame}
Consider Figure 1 of section \ref{subsubsec:thought} with time ($T>0)$ 
in the right side of the figure when the cw light pulse reaches the disk 
observer designated as $T_{1}$. Consider the time when the ccw pulse reaches 
the disk observer (not shown) as $T_{2}$. Then lengths traveled as seen from 
the lab by the ccw light pulse and the observer at $T_{1}$ must sum to equal 
the circumference, i.e.
\begin{equation}
\label{eq25}
cT_1 +\omega RT_1 =2\pi R\quad \quad \to \quad \quad T_1 =\frac{2\pi 
R}{c+\omega R}.
\end{equation}
Similarly, at time $T_{2}$ 
\begin{equation}
\label{eq26}
cT_2 =\omega RT_2 +2\pi R\quad \quad \to \quad \quad T_2 =\frac{2\pi 
R}{c-\omega R}.
\end{equation}
Hence, the arrival time difference in the lab is
\begin{equation}
\label{eq27}
\Delta T=T_2 -T_1 =\frac{2\omega R}{c^2}\frac{2\pi R}{(1-\omega 
^2R^2/c^2)}=\frac{4\omega A}{c^2(1-\omega ^2R^2/c^2)}.
\end{equation}
As is well known, the standard (physical) clocks on the disk rim run more 
slowly than the lab clocks by $\sqrt {1-\omega ^2R^2/c^2} $, so the observer 
on the disk must measure an arrival time difference of
\begin{equation}
\label{eq28}
\Delta t_{phys} =\frac{4\omega A}{c^2\sqrt {1-\omega ^2R^2/c^2} }.
\end{equation}

%%for bibliography, replace usual stuff with

%\begin{chapthebibliography}{<widest bib entry>}
%% they used 99 for widest bib entry
%%then the usual entries as follows (next 2 lines, e.g.)
%%\bibitem{} Authors, \textit{Review}, \textbf{volume}, page (year)
%%\bibitem{} Authors, \textit{Title of The Book/Proceedings}, Publisher, City, (year)
%\end{chapthebibliography}


\begin{thebibliography}{12}
\bibitem{Transforming:1} Transforming from the 4D Riemann flat space of the lab to the 4D space of the rotating frame implies the 4D space of the rotating frame is also Riemann flat. The traditional analysis does not contradict this, but claims the 2D surface of the rotating disk is curved within the flat 4D rotating frame space.
\bibitem{There:1} There may not be a difference between the NTO and SRT predictions for this experiment, and even with sufficient accuracy, the test might not be capable of discerning between the two approaches. Significant analysis would be required to answer this question.
\bibitem{With:1} With the beam aligned north-south, as it was in the experiment, there is definitely no difference between the NTO and SRT predictions. For an east-west alignment, there may, or may not be a difference. Significant analysis would be required to answer this question.
\end{thebibliography}
\end{document}